\documentclass[12pt, hidelinks]{article}

\usepackage{preample}
\usepackage{hyperref}
\usepackage[margin=1in]{geometry}

\title{Credible, Optimal Auctions via Public Broadcast}

\author{Tarun Chitra \thanks{Gauntlet (\href{mailto:tarun@gauntlet.network}{tarun@gauntlet.network})}
\and
Matheus V. X. Ferreira \thanks{University of Virginia(\href{mailto:matheus@virginia.edu}{matheus@virginia.edu})}
\and
Kshitij Kulkarni \thanks{UC Berkeley (\href{mailto:ksk@eecs.berkeley.edu}{ksk@eecs.berkeley.edu})}
}

\date{\today}

\begin{document}

\maketitle

\begin{abstract}
We study auction design in a setting where agents can communicate over a censorship-resistant broadcast channel like the ones we can implement over a public blockchain.
We seek to design credible, strategyproof auctions in a model that differs from the traditional mechanism design framework because communication is not centralized via the auctioneer.
We prove this allows us to design a larger class of credible auctions where the auctioneer has no incentive to be strategic.
Intuitively, a decentralized communication model weakens the auctioneer's adversarial capabilities because they can only inject messages into the communication channel but not delete, delay, or modify the messages from legitimate buyers.
Our main result is a separation in the following sense: we give the first instance of an auction that is credible only if communication is decentralized.
Moreover, we construct the first two-round auction that is credible, strategyproof, and optimal when bidder valuations are $\alpha$-strongly regular, for $\alpha > 0$.
Our result relies on mild assumptions --- namely, the existence of a broadcast channel and cryptographic commitments.
\end{abstract}

\section{Introduction}\label{sec:intro}
Incentive compatibility for buyers is desirable in auctions due to improvements in user experience.
For example, in a second-price auction, if the highest bidder bids \$10 and the second highest bidder bids \$5, the highest bidder wins and pays \$5.
Thus, for any buyer, bidding the maximum they are willing to pay is an optimal strategy, independently of the strategy of others.
This differs from non-incentive compatible auctions, such as first-price auctions, where optimal strategies are a complex balance between demand and the strategy of competing buyers.

Extending incentive compatibility to auctioneers is increasingly becoming a topic of interest in designing auctions within digital marketplaces.
In online settings, it is challenging to audit auctions and verify the identity of participants. Thus, a strategic auctioneer can act simultaneously as the seller and a buyer to deviate from the promised auction.
For example, in the second-price auction above, buyers must trust the auctioneer can commit to implementing the promised auction.
Otherwise, a strategic auctioneer impersonating a buyer can easily leverage their privileged position to submit a bid of \$9, increasing revenue and reducing the buyer's welfare.

{\em Credibility} is a form of incentive compatibility for auctioneers that formalizes the incentives for an auctioneer to follow their promised specifications.
It is desirable for auctioning objects ranging from Non-Fungible Tokens (NFTs) to online advertising because it ensures auction outcomes are auditable.
The US Department of Justice's 2023 antitrust suit against Google~\cite{doj2023} effectively argues that Google's manipulation of ad auctions from the privileged position of auctioneer caused both buyers and users harm.
Allegedly, the lack of market transparency afforded Google ``power to manipulate the quantity of ad inventory and auction dynamics in ways that allowed it to charge advertisers more than it could in a competitive
market''.
Thus, credibility is not only a compelling objective for regulators, but also for sellers that wish to prove their auctions are fair.

Unfortunately, recent work has highlighted challenges in designing mechanisms that are simultaneously incentive-compatible for both sellers and buyers.
The pioneering work of~\cite{akbarpour20} considered a model where the auctioneer can modulate their private communication with buyers to increase their revenue and potentially reduce buyer welfare via a safe deviation from a promised auction.
Informally, a {\em safe deviation} is any auctioneer deviation that passes undetectable by any buyer alone.
An auction is {\em credible} if safe deviations cannot increase the auctioneer's expected revenue.
For example, an auctioneer waiting for the highest bidder to bid $\$10$ and impersonating a false buyer that bids $\$9$ is a profitable, safe deviation from a second-price auction.
Unfortunately, \cite{akbarpour20}~demonstrated that auctioneer credibility could not coexist with buyer incentive compatibility unless the communication complexity is unbounded: they showed that an ascending price auction is the only credible, strategyproof optimal auction.

On the other hand, if one is willing to assume that the auctioneer and buyers are computationally bounded --- and thus cannot break known cryptographic assumptions --- one can get around the theoretical barriers of \cite{akbarpour20}.
Concretely, \cite{ferreira2020credible}~demonstrated that there are cryptographic auctions that are credible, incentive compatible, and have bounded communication complexity if buyer valuations satisfy a regularity condition.
They proposed the (centralized) {\em Deferred Revelation Auctions ($\dra$)}, a two-round auction that is optimal, strategyproof, and credible under the assumption buyer valuations are $\alpha$-strongly regular for any $\alpha \geq 1$.
They also show their auction is not always credible if $\alpha < 1$ and valuations have unbounded support.
This challenges adopting these auctions because they are only credible if the buyer valuations have tails not heavier than the exponential distribution, i.e., $\alpha$-strongly regular for $\alpha \geq 1$ does not contain the Pareto distribution, for example.

In the same line as \cite{ferreira2020credible}, \cite{essaidi2022credible} proposed the Ascending Deferred Revelation Auction (ADRA), which is strategyproof and credible without requiring any assumption on the distributions.
However, ADRA communication complexity is constant on expectation and unbounded in the worst case.
In contrast, we study auctions with bounded communication in the worst case.

All results above consider a centralized communication model where buyers can only exchange messages with the auctioneer.
This assumption is motivated by the scenario where one buyer does not have prior knowledge of the identity of a second buyer.
Unfortunately, if the communication is centralized, the auctioneer can launch a man-in-the-middle-like attack by censoring, injecting, and modifying messages they were supposed to forward to other participants. 
In contrast, our work explores the design of credible auctions when agents can access a broadcast channel where any buyer can broadcast messages to all participants.
This assumption is well-motivated in an auction in a physical environment like a traditional auction house.
Further, this assumption has also become realistic for auctions implemented over a communication network like the Internet due to the proliferation of censorship-resistant public blockchains.
Our main contribution shows a simple change in the communication model (centralized vs. distributed) affects the design of credible auctions.

We summarize our findings as follows:

\begin{itemize}
    \item {\bf Theorem~\ref{thm:strongly-regular}.} Assume buyer values are drawn independently from an $\alpha$-strongly regular distribution for $\alpha > 0$. 
    Then, the deferred revelation auction with public broadcast is credible, strategyproof, and revenue optimal.
    Moreover, modifying the auction so buyers can only communicate privately with the auctioneer makes the resulting auction not credible.
    
    \item {\bf Theorem~\ref{thm:regular}.} There is a $0$-strongly regular buyer valuation that witnesses that deferred revelation auction with public broadcast is not credible.
\end{itemize}

\subsection{Related Work}
We have already reviewed the most relevant prior work in our earlier discussion.
Our model is similar to \cite{akbarpour20} under the additional assumption of cryptographic primitives plus a public broadcast channel.

The security of auctions using cryptography has been extensively studied in the literature ~\cite{rabin2006time, brandt2001cryptographic}. Notably, Yao’s seminal work on multi-party computation~\cite{yao82} was initially motivated by economic applications.
Recent research has revisited the problem of secure auction design, incorporating novel cryptographic tools such as homomorphic encryption and timed encryption techniques~\cite{tyagi2023riggs}
\cite{glaeser2023cicada}.

However, these approaches come with stronger trust assumptions. For instance, multi-party computation assumes that a majority of participants are honest. In contrast, our setting allows the auctioneer to create multiple identities. 
Furthermore, timed encryption, although an intriguing concept, has seen limited practical applications due to its reliance on stronger cryptographic assumptions. 
Importantly, our goal of reducing the number of auction rounds aims to enhance auction speed, whereas timed encryption would counter this objective by increasing the auction duration.

Credible mechanism design has applications beyond auctions such as in the design of manipulation-resistant decentralized exchanges~\cite{xavier2023credible}, blockchain transaction fee mechanisms~\cite{ganesh2024revisiting, ferreira2021dynamic}, and in Bayesian persuasion~\cite{lin2024credible}.

\subsection{Technical overview}

\cite{akbarpour20} does not consider the existence of a broadcast channel in their framework because they envision auctions executing over the Internet (or over the telephone) and assume buyers do not know the identity of each other beforehand.
Implementing a broadcast channel in this scenario is challenging and draws from years of research in consensus and cryptography, starting from the Byzantine general's problem of ~\cite{lamport2019byzantine}.
This line of research culminated with the Bitcoin blockchain, which provides censorship-resistant consensus at large scale~\cite{nakamoto2008bitcoin}. In the framework of \cite{akbarpour20}, the auctioneer promises to implement an auction and is the nexus of communication with buyers.
A buyer privately sends messages to the auctioneer and trusts that the auctioneer will forward those messages to other buyers.

We propose a simple modification to this framework that, surprisingly, increases the incentive for the auctioneer to commit to a promised auction.
Concretely, rather than sending messages privately to the auctioneer, we assume any agent can broadcast messages.
Once an agent broadcasts a message $m$, all other participants simultaneously learn about message $m$.

Under the new framework, our main contribution is the {\em deferred revelation auction with public broadcast}.
It is similar to the centralized deferred revelation auction of \cite{ferreira2020credible} with the main difference that buyers can now broadcast messages.
Our main result shows $\dra$ with public broadcast is a credible auction, assuming buyer valuations are $\alpha$-strongly regular for all $\alpha > 0$.
Recall \cite{ferreira2020credible} showed centralized $\dra$ is not credible for these buyer valuations.
This has significant practical implications because it provides the first design of a communication-efficient, credible, strategyproof, optimal auction when buyer value distributions have tails as heavy as Pareto distributions.

Informally, the deferred revelation auction with public broadcast is a two-phase auction (see \S\ref{sec:dra}) as follows:

\begin{enumerate}
\item In the {\em bidding phase}, each buyer broadcasts a cryptographic commitment of their bid and deposits a collateral.

\item The auctioneer broadcasts the end of the commitment phase and the start of the revelation phase.

\item In the {\em revelation phase}, each buyer broadcasts the opening of their commitment (e.g., their bid and the random seed used to generate the commitment).

\item The auctioneer marks a bid as revealed if the second phase message opens the cryptographic commitment received in the first phase.
Then, the auctioneer implements the second-price auction with reserves using the revealed bids.

\item The auctioneer refunds the collateral if, and only if, a buyer reveals their bid.
The confiscated collateral is given to the winner of the auction.
\end{enumerate}

As in~\cite{ferreira2020credible}, we consider a threat model where the auctioneer can shill bid (i.e., impersonate false buyers that submit false bids).
To argue the credibility of our auction, we must show that under certain conditions, sufficiently large collateral incentivizes the auctioneer not to impersonate false buyers.
Central to our argument is observing that the security of our cryptographic commitment scheme (see Definition~\ref{def:commitment}) together with a broadcast channel ensures the auctioneer cannot commit to a bid that depends on the bids of other buyers.
This is not the case for the centralized deferred revelation auction.
To see, consider modifying the auction above so that whenever a buyer broadcasts a message, the buyer sends that message to the auctioneer, who ``promises'' to forward it to all other buyers.
The following is a safe deviation to centralized $\dra$ where shill bids depends on bids from genuine buyers.

\begin{example}
Suppose there are genuine buyers $A$ and $B$ as well as a false buyer $C$.
Any message the auctioneer receives from $B$, the auctioneer forwards to $A$.
The auctioneer does not forward any message from $A$ to $B$ which makes buyer $B$ unaware that $A$ exists.
The auctioneer asks buyer $A$ to open their bid and after learning the bid $b_A$ of $A$, the auctioneer impersonates a false buyer $C$ that commits to bid $b_A + \Delta$ to buyer $B$.
This deviation is undetectable because buyer $A$ believes only $A$ and $B$ participate in the auction.
Moreover, $A$ cannot detect their messages were censored.
On the other hand, $B$ believes only $B$ and $C$ participate in the auction.
Moreover, $B$ cannot detect that $A$'s messages were censored (in fact, $B$ is unaware of $A$).
Finally, observe that $B$ receives a bid from a false buyer correlated with the bid of $A$.
\end{example}

This might seem like an innocent deviation, but  Section~\ref{sec:necessary} shows centralized $\dra$ is not credible for $\alpha$-strongly regular valuations for any $\alpha \in [0, 1)$ if we adapt this strategy and allow the auctioneer to submit shill bids that depend on genuine bids.
Clearly this deviation is not possible if a broadcast channel is available since the auctioneer cannot choose who gets to observe $A$'s messages and the auctioneer cannot commit to shill bids after starting the revelation phase.
Our main technical contribution shows that safe deviations that leverage shill bids correlated to genuine bids were the only strategies that prevented centralized $\dra$ from being a credible auction when buyer valuations are $\alpha$-strongly regular for $\alpha > 0$.

\subsection{Paper organization}
We provide the necessary background in optimal auction theory in \S\ref{sec:preliminaries}. We define the implementation of the deferred revelation auction with public broadcasts in \S\ref{sec:dra}.
We prove our main result, Theorem~\ref{thm:strongly-regular}, in \S\ref{sec:dra-credible} and our negative result, Theorem~\ref{thm:regular}, in \S\ref{sec:regular}.
\S\ref{sec:necessary} shows that a broadcast channel is necessary for Theorem~\ref{thm:strongly-regular}.
We conclude in \S\ref{sec:conclusion} and include future directions.

\section{Preliminaries}\label{sec:preliminaries}
We consider a single item, $n$ buyer auction. Buyer $i \in [n] = \{1, \ldots, n\}$ has private value $v_i \in \mathbb R^+$ and has quasilinear utility: if they receive the item and pay $p_i$, then their utility is $v_i - p_i$; if they do not receive the item, then their utility is $0$. We assume $v_i$ is drawn independently from a distribution $D$ with CDF $F$ and PDF $f$. The auctioneer knows the distribution $D$, but not the values $\{v_i\}_{i = 1}^n$. We refer to $\vec v = (v_1, \ldots, v_n)$ as a {\em value profile}. We write the value profile of all buyers except buyer $i$ as $\vec v_{-i} = (v_1, \ldots, v_{i-1}, v_{i+1}, \ldots, v_n)$.

\vspace{1mm} \noindent {\bf Communication model.} Agents can communicate on a {\em private channel} or a {\em broadcast channel}. If agent $i$ sends a message $m$ in a broadcast channel, then the message is immediately received buy all other agents. If agent $i$ sends a message $m$ to agent $j \neq i$ in a private channel, only agent $j$ observes $m$.

\vspace{1mm} \noindent {\bf Extensive-form game.} An extensive-form game $G$ consists of a tree $(H, E)$ where the nodes $H$ are the set of histories and edges $E \subseteq H \times H$ are state transitions. The game starts at the root of $(H, E)$, has a set of players $\{0, 1, \ldots, n\}$, and a collection of actions $A(h)$ available at each history $h \in H$. We refer to player $0$ as the auctioneer and player $i \in [n]$ as buyer $i$. Each history $h \in H$ has one owner $P(h) \in \{0, \ldots, n\}$ responsible for taking the next action when the game is at state $h$. After taking action $a \in A(h)$, the game moves to another history $h'$ where $(h, h') \in E$. We consider games of incomplete information where only agent $P(h)$ observes the action $A(h)$ taken at $h$. 

A strategy $s_i$ for buyer $i \in [n]$ on game $G$ is a function that takes buyer $i$'s private type $v_i$ and any history $h \in H$ where $i \in P(h)$ and outputs the agent's action $s_i(v_i, h) \in A(h)$ at $h$. Consider a strategy profile $\vec s = (s_1, \ldots, s_n)$. An {\em auction game} $(G, \vec s)$ is a communication game on $G$ when buyers follow strategy $\vec s$ that allocates the item and charges payments.

The outcome of auction game $(G, \vec s)$ is a tuple $(\vec x^{(G, \vec s)}(\vec v), \vec p^{(G, \vec s)}(\vec v))$ where $x_i^{(G, \vec s)}(v)$ and $p_i^{(G, \vec s)}(\vec v)$ denotes the probability that agent $i$ receives the item and their payment respectively. A strategy $s_i$ is a best response to $\vec s_{-i}$ if for any strategy $s_i'$ for buyer $i$, for any $\vec v$,
$$v_i \cdot x_i^{(G, \vec s)}(\vec v) - p_i^{(G, \vec s)}(\vec v) \geq v_i \cdot x_i^{(G, s_i', \vec s_{-i})}(\vec v) - p_i^{(G, s_i', \vec s_{-i})}(\vec v).$$

\begin{definition}[Ex-post Nash/Strategyproof/Individually Rational]
Consider an auction $(G, \vec s)$. A strategy profile $\vec s$ forms an {\em ex-post Nash equilibrium}, if for any buyer $i$, strategy $s_i$ is the best response to $\vec s_{-i}$. An auction is {\em strategyproof} if some strategy profile $\vec s$ forms an ex-post Nash equilibrium. An auction is {\em individually rational (IR)} if there is a strategy for any buyer that ensures such buyer receives non-negative utility.
\end{definition}

The auctioneer's {\em expected revenue} on auction game $(G, \vec s)$ is $\rev(G, \vec s) := \e[\vec v]{\sum_{i = 1}^n p_i^{(G, \vec s)}(\vec v)}$.

We assume the auctioneer can deviate from implementing $(G, \vec s)$ as long as any buyer cannot detect a deviation. These are a {\em safe deviation} from the promised auction $(G, \vec s)$. Formally, $(G', s)$ is a {\em safe deviation} from $(G, \vec s)$ if for any buyer $i \in [n]$, there is a strategy profile $s_{-i}^i = (s_1^i, \ldots, s_{i-1}^i, s_{i+1}^i, \ldots, s_{n_i}^i)$ for $n_i$ buyers where buyer $i$ observes the same messages in communication games $(G', \vec s)$ and $(G, s_i, s_{-i}^i)$.

\begin{definition}[Credible Auction]
An auction game $(G, \vec s)$ is {\em credible} if for any safe deviation $(G', \vec{s})$ of $(G, \vec s)$, $\rev(G, \vec s) \geq \rev(G', \vec s)$.
\end{definition}

\subsection{Virtual values} Virtual values functions allow us to formalize optimal auctions. The {\em virtual value function} associated with continuous CDF $F$ and PDF $f$ is $\varphi^F(x) = x - \frac{1-F(x)}{f(x)}$. We write $\varphi(\cdot)$, omitting the superscript $F$, when the distribution is clear from the context. We write $\varphi^+(x) = \max\{0, \varphi(x)\}$. A distribution $F$ is {\em $\alpha$-strongly regular} for $\alpha \geq 0$ if for all $x' \geq x$,
$$\varphi(x') - \varphi(x) \geq \alpha (x' - x).$$
A distribution $F$ has {\em Monotone Hazard Rate (MHR)}, if $F$ is $1$-strongly regular. A distribution is {\em regular} if $F$ is $0$-strongly regular.
Note that MHR distributions have exponentially decaying tails, whereas distributions with $\alpha \in (0,1)$ have polynomially decaying tails.

\begin{theorem}[Myerson's Theorem~\cite{myerson1981optimal}]\label{lemma:myerson}\label{thm:myerson}
Consider a strategyproof auction that awards the item to buyer $i$ with probability $x_i(\vec v)$ and charges $p_i(\vec v)$ on bids $\vec v$. Then, the expected revenue is
$$\e[\vec v \leftarrow D]{\sum_{i = 1}^n p_i(\vec v)} = \e[\vec v \leftarrow D]{\sum_{i = 1}^n \varphi(v_i) \cdot x_i(\vec v)}.$$
We refer to the right-hand side as the expected {\em virtual welfare}. For cases where $D$ is regular, $\varphi$ is non-decreasing, and the optimal auction maximizes expected virtual welfare.
\end{theorem}
We define the inverse of a monotone function $g(\cdot)$ to be $g^{-1}(y)=\inf_x \{x|\ g(x) \geq y\}$. We define to $r(D) := (\varphi^D)^{-1}(0)$ as {\em Myerson's reserve price}. From Myerson's theorem, the optimal auction only sells the item to buyers with the value $v_i \geq r(D)$. We define $\rev(D^n)$ as the expected revenue of the optimal auction with $n$ buyers with valuations drawn i.i.d. from $D$. We provide facts about $\alpha$-strongly regular distributions in Appendix~\ref{app:background}.

\section{Deferred Revelation Auction (DRA) with Public Broadcast}\label{sec:dra}
This section defines the deferred revelation auction with public broadcast. The central assumption is the existence of a perfectly hiding, computationally binding, and non-malleable cryptographic commitment scheme as follows.

\begin{definition}[Commitment Scheme]\label{def:commitment} A {\em commitment scheme} is a function $\commit(\cdot, \cdot)$ that takes a message $m \in \{0, 1\}^*$, a random string $r \in \{0, 1\}^\lambda$ where $\lambda \in \mathbb N$ is the {\em security parameter} and outputs a commitment $c \in \{0, 1\}^{\poly(\lambda)}$ where $\poly(\lambda)$ is a polynomial with variable $\lambda$.

\vspace{1mm} {\noindent \bf Perfectly hiding.} A commitment scheme is {\em perfectly hiding} if, for all $m \neq m'$, $\commit(m, r)$ is identically distributed to $\commit(m', r')$ provided that $r$ and $r'$ are uniformly random.

\vspace{1mm} {\noindent \bf Computationally binding.} A commitment scheme is {\em computationally binding} if for any probabilistic polynomial time algorithm that takes the security parameter $\lambda$ and terminates in expected time $\poly(\lambda)$, then the probability the algorithm outputs $(m, r) \neq (m', r')$ such that $\commit(m, r) = \commit(m', r')$ is at most $2^{-\lambda}$.

\vspace{1mm} {\noindent \bf Non-malleable.} Consider any communication game where a probabilistic polynomial time adversary receives $c = \commit(m, r)$ where $m$ is drawn from a known distribution and $r$ is uniformly random. In the first round, the adversary must output some commitment $c' \neq c$. In the second round, the attacker learns $(m, r)$ 
and outputs $(m', r')$ such that $\commit(m', r') = c$.
We say the commitment scheme is {\em non-malleable} if, for any such game, the random variable $(m', r')$ is independent of $(m, r)$.
\end{definition}

Some commitment schemes are malleable; for example, they allow a receiver that observes $\commit(b, r)$ to compute $\commit(b-1, r)$.
This does not violate secrecy since the receiver does not learn $b$ or can open $\commit(b-1, r)$ before the sender opens $(b, r)$).
Yet, this malleability would pose serious security vulnerabilities in an auction.
If a bidder commits to bid $b$ with $\commit(b, r)$, the auctioneer can shill bid and commit to bidding $b-1$ by computing $\commit(b-1, r)$.
Constructions of non-malleable commitment schemes are involved and outside the scope of this work (see \cite{khurana2017achieve, fischlin2000efficient} for a more general definition and practical constructions).

\begin{definition}[Deferred Revelation Auction with Public Broadcast] Let $\commit(\cdot,\cdot)$ be a perfectly hiding, perfectly binding, and non-malleable commitment scheme satisfying Definition~\ref{def:commitment}. A {\em collateral function} $f(\cdot,\cdot)$ takes the number of buyers $n$ and a distribution $D$ and outputs a collateral required from each buyer to bid in the auction. For a collateral function $f$, {\em $\dra(f)$ with public broadcast} is the following auction:

\paragraph{Commitment phase ($1^{st}$ round):}
\begin{itemize}
    \item Each buyer $i \in [n]$ picks a bid $b_i = v_i$, draws $r_i$ uniformly at random, and broadcast $(i, \commit(b_i,r_i))$. Moreover, buyer $i$ sends collateral $f(n, D)$ to the auctioneer.

    \item The auctioneer broadcasts ``End of Commitment Phase".
\end{itemize}

\paragraph{Revelation phase ($2^{nd}$ round):}
\begin{itemize}
    \item Each buyer $i$ broadcasts $(i, b_i', r_i')$ where $b_i' = b_i$ and $r_i' = r_i$.

    \item The auctioneer broadcasts ``End of Revelation Phase".
\end{itemize}

\paragraph{Resolution phase:}
\begin{itemize}
\item Let $S$ denote the set of buyers for which $\commit(b_i, r_i) = \commit(b_i', r_i')$. Let $b'_i := b_i \cdot \ind{i \in S}$. Let $i^* := \arg\max_{i \in S} b_i$.
    
\item If $b_{i^*} > r(D)$, award buyer $i^*$ the item. Charge them
$$\max\{r(D), \max_{i \in S \setminus \{i^*\}} b_i\}.$$

\item The auctioneer refunds the collateral of buyer $i \in S$.

\item The auctioneer transfers the collateral of each buyer $i \not\in S$ to buyer $i^*$.
\end{itemize}

\paragraph{Tie-breaking:}
\begin{itemize}
    \item All ties are broken lexicographically, with the auctioneer treated as ``buyer zero''.
\end{itemize}
\end{definition}

Before discussing how our auction differs from centralized $\dra(f)$, we quickly observe that $\dra(f)$ with public broadcast is indeed strategyproof and revenue optimal.

\begin{theorem}\label{thm:dra-optimal}
For all $f$, $\dra(f)$ with public broadcast is a strategyproof optimal auction.
\end{theorem}
\begin{proof}
The definition for $\dra(f)$ instructs each buyer $i \in [n]$ to follow the strategy where buyer $i$ sets $b_i = v_i$; in the commitment phase, buyer $i$ picks a uniformly random $r_i$ and broadcasts a commitment $\commit(v_i, r_i)$; and in the revelation phase, buyer $i$ reveals $(v_i, r_i)$.
Since $\dra(f)$ implements the same outcome as a second-price auction, it follows this strategy profile and is an ex-post Nash equilibrium, which proves the auction is strategyproof. 
Moreover, because the auction maximizes expected virtual welfare,  Theorem~\ref{thm:myerson} (Myerson's Theorem) implies the auction is revenue optimal.
\end{proof}

Definition~\ref{def:centralized-dra} provides a definition for centralized $\dra(f)$ \cite{ferreira2020credible}. 
Lemma~\ref{lemma:strategy-space} shows that centralized $\dra(f)$ has strategy space for the auctioneer at least as ample as $\dra(f)$ with public broadcast. To be concrete, the lemma shows that any safe deviation to $\dra(f)$ with public broadcast maps to a safe deviation to centralized $\dra(f)$.

\begin{definition}[Centralized Deferred Revelation Auction]\label{def:centralized-dra}
    The {\em centralized $\dra(f)$} is identical to $\dra(f)$ with public broadcast except under the following cases:
    \begin{itemize}
        \item In $\dra(f)$ with public broadcast, consider a history $h$ where buyer $i$ broadcasts a message $m$.
        In centralized $\dra(f)$, instead of broadcasting $m$, buyer $i$ sends $m$ to the auctioneer in a private channel.
        Then, the auctioneer sends $m$ to each buyer $j \neq i$ in a private channel.

        \item In $\dra(f)$ with public broadcast, consider a history $h$ where the auctioneer broadcasts a message $m$.
        In centralized $\dra(f)$, instead of broadcasting $m$, the auctioneer sends $m$ to each buyer $i \in [n]$ in a private channel.
    \end{itemize}
\end{definition}

\begin{lemma}\label{lemma:strategy-space}
Let $(G, \vec s)$ be a safe deviation to $\dra(f)$ with public broadcast, then there is a safe deviation $(G', \vec s')$ to centralized $\dra(f)$ where $\rev(G', \vec s') = \rev(G, s)$
\end{lemma}
\begin{proof}
Let $(G', \vec s')$ be a communication game identical to $(G, \vec s)$ except on the following scenario:

\begin{itemize}
    \item Whenever buyer $i \in [n]$ broadcasts message $m$ in $(G, \vec s)$, in $(G', \vec s')$, buyer $i$ sends $m$ to the auctioneer.
    After receiving $m$, the auctioneer sends $m$ to each buyer $j \neq i$.

    \item Whenever the auctioneer broadcasts message $m$ in $(G, \vec s)$, in $(G', \vec s')$, the auctioneer sends $m$ to each buyer $i \in [n]$.
\end{itemize}

The deviation $(G', \vec s')$ is safe assuming $(G, \vec s)$ is safe. Moreover, it induces the same allocation/payment rules, meaning it obtains the same revenue as $(G, \vec s)$. This concludes the proof.
\end{proof}

Unfortunately, the converse of Lemma~\ref{lemma:strategy-space} is untrue. There are safe deviations to centralized $\dra(f)$ that do not map to any safe deviation in $\dra(f)$ with public broadcast. We give the following examples to illustrate this fact.
\begin{example}
In $\dra(f)$ with public broadcast, buyer $i$ sends $(i, c_i)$ to all buyers. On the other hand, in centralized $\dra(f)$, buyer $i$ must send $(i, c_i)$ to the auctioneer, and the auctioneer ``promises'' to forward $(i, c_i)$ to all buyers $j \neq i$. Unfortunately, buyer $i$ cannot verify whether the auctioneer forwards their message to any buyer $j \neq i$. This allows the auctioneer to share $(i, c_i)$ with a strict subset of buyers.
\end{example}

\begin{example}
In $\dra(f)$ with public broadcast, the auctioneer broadcasts the end of the commitment phase to all buyers. On the other hand, on centralized $\dra(f)$, the auctioneer ``promises'' to simultaneously announce the end of the commitment for each buyer. Suppose the auctioneer announces the end of the commitment phase to buyer $i$ at 10:00 p.m. but only sends this announcement to buyer $j$ at 11:00 p.m. This deviation is safe because buyer $i$ does not know which messages buyer $j$ received and vice-versa. Thus, at 10:10 p.m., the auctioneer requests buyer $i$ to reveal $(b_i, r_i)$. Then, the auctioneer impersonates a false buyer $z$ that bids $b_z(b_i)$ that might depend on $b_i$. Buyer $z$ sends $\commit(b_z(b_i), r_z)$ only to buyer $j$ at 10:20 p.m.
\end{example}

These examples do not prove there are safe deviations to centralized $\dra(f)$ that are more profitable than any safe deviation to $\dra(f)$ with public broadcast.
They aim to showcase additional manipulations the auctioneer can perform that they cannot perform when a broadcast channel is present.
Our main result will formally prove that these manipulations strictly improve the auctioneer's revenue relative to deviations that do not manipulate the order and time of messages.

Note some deviations are still possible even when buyers communicate in a broadcast channel, which makes arguing about the credibility of $\dra(f)$ with public broadcast non-trivial --- namely, the addition of false bids and the refusal to reveal false bids.

\vspace{1mm} \noindent {\bf Broadcasting false bids.} In the commitment phase, the auctioneer can impersonate a {\em false buyer} --- agents that submit bids not coming from any {\em real buyer} $i \in [n]$ --- which broadcast a {\em false bid} $\commit(\hat b, \hat r)$ where $\hat r$ is uniformly random. We refer to $\tilde b(n, D)$ as the highest bid among all false buyers. Set $\tilde b(n, D) = 0$ if the auctioneer does not impersonate any false buyer.

\begin{lemma}\label{lemma:false-bid-independence}
Assume the auctioneer follows a safe deviation to $\dra(f)$ with public broadcast. If, during the commitment phase, a false buyer broadcasts $\commit(b, r)$, and, in the revelation phase, the false buyer reveals $(b, r)$, then $b$ is a random variable independent of $\vec v$.
\end{lemma}
\begin{proof}
Suppose for contradiction the false buyer broadcasts $\commit(b, r)$ and later reveals $(b, r)$ where $b$ is not independent of $\vec v$. Use this auction to construct an adversary that outputs $\commit(b, r)$ whenever the false buyer does. Once the false buyer reveals $(b, r)$, the adversary reveals $(b, r)$. Because $b$ is correlated to $\vec v$, this implies the commitment scheme is malleable, a contradiction.
\end{proof}

\vspace{1mm} \noindent {\bf Withhold false bids.} In the revelation phase, the auctioneer can refuse to reveal any bid $\hat b$ submitted from a false buyer.
The decision to reveal or withhold a bid from a false buyer might depend on the real bids $\vec b$.

Next, we highlight a few relevant facts about our protocol. In the commitment phase, buyer $i$ observes commitments $\{\commit(d_j^i, r_j^i)\}_j$ from both real buyers and false buyers (excluding their bid $b_i$). That is, $d_j^i$ is the $j$-th bid buyer $i$ observes excluding their own bid. Let $\beta_i(\vec b) = \max\{r(D), \max_j \{d_j^i\}\}$ be the highest bid buyer $i$ observed in the commitment phase (including the reserve price $r(D)$ and excluding their bid $b_i$) when real buyers bid $\vec b$. It is possible $\max_{i \in [n]} \beta_i(\vec b) > \max\{r(D), \max_{i \in [n]}\{b_i\}\}$ if the highest bid is from a false buyer.

\begin{observation}\label{obs:single-candidate}
Assume the auctioneer follows a safe deviation to $\dra(f)$ with public broadcast. Then for all value profiles $\vec b$, $b_i > \beta_i(\vec b)$ for at most one buyer $i \in [n]$.
\end{observation}
\begin{proof}
    Suppose for contradiction there are distinct buyers $i$ and $j$ such that $b_i > \beta_i(\vec b)$ and $b_j > \beta_j(\vec b)$. Observe that buyer $i$ receives the bid of buyer $j$ and buyer $j$ receives the bid of buyer $i$ which implies $\beta_i(\vec b) \geq b_j$ and $\beta_j(\vec b) \geq b_i$. The inequalities implies $b_i > b_j$ and $b_j > b_i$, a contradiction. This proves there is at most one buyer $i$ such that $b_i > \beta_i(\vec b)$.
\end{proof}

\begin{observation}\label{obs:safe-allocation}
Suppose the auctioneer follows a safe deviation to $\dra(f)$ with public broadcast. If $b_i > \beta_i(\vec b)$, then buyer $i$ receives the item and pays $\beta_i(\vec b)$.
\end{observation}
\begin{proof}
Buyer $i$ can observe that their bid is above the reserve price and they are the highest bidder in the auction. If the auctioneer's deviation is safe, it must allocate the item to $b_i$ and charge $\beta_i(\vec b)$.
\end{proof}

The following Lemma~\ref{lemma:collateral} shows that under certain conditions, it is optimal for the auctioneer to reveal any bids from false buyers. 
\begin{lemma}\label{lemma:collateral}
Consider any safe deviation $(G, \vec s)$ of $DRA(f)$ where, in the commitment phase, the auctioneer impersonates a false buyer that bids $0 < b \leq f(n, D)$, and, in the revelation phase, the auctioneer withholds $b$.
Let $h$ be the history where the auctioneer reveals or withholds $b$.
Let $(G', \vec s)$ be a new deviation identical to $(G, \vec s)$ except at history $h$ the auctioneer reveals $b$.
Then $G'$ is a safe deviation and $\rev(G', \vec s) \geq \rev(G, \vec s)$.
\end{lemma}
\begin{proof}
The fact $(G', \vec s)$ is a safe deviation follows directly from the fact $(G, \vec s)$ is a safe deviation.
Next, we argue $\rev(G', \vec s) \geq \rev(G, \vec s)$.

First, consider the case where no real buyer receives the item at $(G, \vec s)$.
Then, no real buyer will receive the item at $(G', \vec s)$.
Moreover, $(G', \vec s)$ improves the auctioneer's revenue relative to $(G, \vec s)$ because the auctioneer receives no payments but pays fewer penalties for revealing $b$.

Next, consider the case where some buyer $i \in [n]$ receives the item and pays $p$ at $(G, \vec s)$.
For the case where $b \leq p$, buyer $i$ remains the highest bidder and pays $p$ while the auctioneer pays fewer penalties for revealing $b$ at $(G', \vec s)$.
For the case where $b > p$, by assumption $f(n, D) \geq b$.
Then, the auctioneer receives negative profits at $(G, \vec s)$ since the penalty for withholding $b$ is higher than the payment they receive from buyer $i$.
On the other hand, at $(G', \vec s)$, the revenue loss for revealing $b$ is lower than the penalties for withholding $b$.
\end{proof}

\section{DRA with Public Broadcast is Credible for $\alpha$-Strongly Regular Distributions}\label{sec:dra-credible}
In this section, we show that for any $\alpha$-strongly regular distributions for $\alpha > 0$, there is a $f(\cdot, \cdot)$ that makes $\dra(f)$ with public broadcast credible. Recall that $\tilde b(n, D)$ is the most significant bid from a false buyer. From Lemma~\ref{lemma:false-bid-independence} $\tilde b(n, D)$ is independent of $\vec v$.

For the case where $\alpha \geq 1$, \cite{ferreira2020credible} proved Theorem~\ref{thm:centralized-dra} stating centralized $\dra(f)$ is a credible auction if we set the collateral to be at least the optimal reserve price. Extending their result for our auction is a simple observation that any safe deviation for $\dra(f)$ with public broadcast is also a safe deviation for centralized $\dra(f)$. 

\begin{theorem}[Theorem 4.1 in \cite{ferreira2020credible}]\label{thm:centralized-dra}
Assume buyer valuations are $\alpha$-strongly regular for any $\alpha \geq 1$. If $f(n, D) \geq r(D)$, then centralized $\dra(f)$ is a credible auction.
\end{theorem}

\begin{theorem}\label{thm:ruduction-centralized-to-broadcast}
Assume buyer valuations are $\alpha$-strongly regular for any $\alpha \geq 1$. If $f(n, D) \geq r(D)$, then $\dra(f)$ with public broadcast is a credible auction.
\end{theorem}
\begin{proof}
Suppose for contradiction $\dra(f)$ with public broadcast is not a credible auction when $f(n, D) \geq r(D)$. There is a safe deviation $(G, \vec s)$ to $\dra(f)$ with public broadcast where $\rev(G, \vec s) > \rev(D^n)$. From Lemma~\ref{lemma:strategy-space}, there is a safe deviation $(G', \vec s')$ to centralized $\dra(f)$ where $\rev(G', \vec s') = \rev(G, s) > \rev(D^n)$. Thus, centralized $\dra(f)$ is not a credible auction, a contradiction to Theorem~\ref{thm:centralized-dra}.
\end{proof}

The challenging case is to argue $\dra(f)$ with public broadcast is credible for some $f(n, D)$ for the case where $\alpha \in (0, 1)$. We first show that any safe deviation where false buyers only broadcast bids smaller than the collateral cannot improve the auctioneer's revenue.

\begin{lemma}\label{lemma:high-collateral}
Assume the auctioneer follows a safe deviation to $\dra(f)$ with public broadcast. Let $\tilde b(n, D)$ be the highest bid from a false buyer (or zero if there are no false buyers). If $f(n, D) \geq \tilde b(n, D)$, the auctioneer's revenue is at most $\rev(D^n)$.
\end{lemma}
\begin{proof}
From Lemma~\ref{lemma:collateral} and the fact the highest false buyer bids $k$, it is without loss of generality to assume the auctioneer always will reveal $\tilde b(n, D)$.

Suppose the reserve price is $r$ and let $\hat r = \max\{r(D), \tilde b(n, D)\}$. Note that $\hat r$ is independent of $\vec v$ because $r(D)$ depends only on $D$ and $b(n, D)$ depends only on $n$ and $D$. Thus, the allocation/payment rule is equivalent to a second-price auction with reserve $\hat r$. Since the second-price auction with reserve $\hat r$ is a strategyproof auction, Myerson's theorem implies the revenue is at most:
$$\e[v \leftarrow D]{\sum_{i = 1}^n p_i(\vec v)} = \e[v \leftarrow D]{\sum_{i = 1}^n \varphi(v_i) \cdot x_i(\vec v)} \leq \e{\max_i \varphi^+(v_i)}.$$
The first equality is Theorem~\ref{thm:myerson}. The second inequality observes $\sum_{i = 1}^n x_i(\vec v) \leq 1$. From Myerson's theorem, the optimal auction maximizes virtual surplus or equivalently, $\rev(D^n) = \e{\max_i \varphi(v_i)}$. This concludes the proof.
\end{proof}

Next, we consider the case where false buyers might broadcast bids higher than the collateral. Our first Lemma will bound the revenue for events where $v_j > \beta_j(\vec v)$ for some buyer $j$. The second Lemma bounds the revenue for events where $v_j < \beta_j(\vec v)$ for all buyers.

\begin{lemma}\label{lemma:safe-allocation} Assume the auctioneer follows a safe deviation to $\dra(f)$ with public broadcast. Let $R(\vec v)$ be the auctioneer's revenue when buyers have value profile $\vec v$. Then
$$\e[\vec v \leftarrow D]{R(\vec v) \cdot \ind{\exists j, v_j > \beta_j(\vec v)}} \leq \e[\vec v \leftarrow D]{\sum_{i = 1}^n \varphi(v_i) \cdot \ind{v_i > \beta_i(\vec v)}}$$
\end{lemma}
\begin{proof}
From Observation~\ref{obs:single-candidate}, there is at most one buyer $i$ such that $v_i > \beta_i(\vec v)$ for any $\vec v$.
Moreover, when $v_i > \beta_i(\vec v)$, buyer $i$ wins the item and pay $\beta_i(\vec v)$.
Since $\beta_i(\vec v)$ is independent of $v_i$, this payment/allocation rule is strategyproof.
From Myerson's theorem, the revenue is the expected virtual surplus $\e[\vec v \leftarrow D]{\varphi(v_i) \cdot \ind{v_i > \beta_i(\vec v)}}$.
We obtain
\begin{align*}
&\e[\vec v \leftarrow D]{R(\vec v) \cdot \ind{\exists j, v_j > \beta_j(\vec v)}}
= \e[\vec v \leftarrow D]{\sum_{i = 1}^n \beta_i(\vec v) \cdot \ind{v_i > \beta_i(\vec v)}}\\
&\qquad = \e[\vec v \leftarrow D]{\sum_{i = 1}^n \varphi(v_i) \cdot \ind{v_i > \beta_i(\vec v)}} \qquad \text{\{By Theorem~\ref{thm:myerson}\}}
\end{align*}
as desired.
\end{proof}

\begin{lemma}\label{lemma:small-collateral}
Assume the auctioneer follows a safe deviation to $\dra(f)$ with public broadcast. Assume $D$ is $\alpha$-strongly regular for $\alpha \in (0, 1)$. Let $\tilde b(n, D)$ be the highest bid from a false buyer (or zero if there are no false buyers). Assume $f(n, D) \geq r(D) \left(\frac{n}{\alpha}\right)^{\frac{1-\alpha}{\alpha}}\left(\frac{1}{1-\alpha}\right)^{\frac{1}{\alpha}}$ and $\tilde b(n, D) > f(n, D)$. Let $R(\vec v)$ be the auctioneer's revenue when buyers have value profile $\vec v$. Then, the auctioneer's expected revenue is at most 
$$\e[\vec v \leftarrow D]{R(\vec v) \cdot \ind{\forall j, v_j < \beta_j}} \leq \rev(D^n) - \e[\vec v \leftarrow D]{\sum_{i = 1}^n \varphi(v_i) \cdot \ind{v_i > \beta_i(\vec v)}}.$$
\end{lemma}

\begin{proof}
When $v_j < \beta_j(\vec v)$ for all buyers, a false buyer is the highest bidder. Therefore, $\max_j \beta_j(\vec v) = \tilde b(n, D) > f(n, D)$ where the inequality is a statement assumption.
In this case, any buyer $j$ can receive the item as long as the auctioneer withholds at least one bid.
Because buyer $j$ pays at most $v_j$, the auctioneer receives negative revenue if $v_j < f(n, D)$. Recall $x_i(\vec v)$ is an indicator variable taking value $1$ if and only if buyer $i$ receives the item. This gives
\begin{align*}
& \e[\vec v \leftarrow D]{R(\vec v) \cdot \ind{\forall j, v_j < \beta_j(\vec v)}}\\
&\qquad\leq \e[v \leftarrow D]{\sum_{i = 1}^n (v_i - f(n, D)) \cdot x_i(\vec v) \cdot \ind{\forall j, v_j < \beta_j(\vec v)} \cdot \ind{v_i \geq f(n, D)}}\\
&\qquad\leq \e[\vec v \leftarrow D]{\sum_{i = 1}^n \left(\frac{1}{\alpha}\varphi(v_i) + r(D) - f(n, D)\right) \cdot x_i(\vec v) \cdot \ind{\forall j, v_j < \beta_j(\vec v)} \cdot \ind{v_i \geq f(n, D)}}\\
&\qquad\leq \e[\vec v \leftarrow D]{\sum_{i = 1}^n \frac{1}{\alpha}\varphi(v_i) \cdot x_i(\vec v) \cdot \ind{\forall j , v_j < \beta_j(\vec v)} \cdot \ind{v_i \geq f(n, D)}}\\
&\qquad= \e[\vec v \leftarrow D]{\sum_{i = 1}^n \frac{1}{\alpha}\varphi(v_i) \cdot x_i(\vec v) \cdot \ind{\forall j \neq i, v_j < \beta_j(\vec v)} \cdot \ind{f(n, D) \leq v_i < \beta_i(\vec v)}}\\
&\qquad= \e[\vec v \leftarrow D]{\sum_{i = 1}^n \frac{\varphi(v_i)}{\alpha} \cdot x_i(\vec v) \cdot \ind{\forall j \neq i, v_j < \beta_j(\vec v)} \cdot (\ind{v_i \geq f(n, D)} - \ind{v_i > \beta_i(\vec v)})}\\
&\qquad < \e[\vec v \leftarrow D]{\sum_{i = 1}^n \frac{\varphi(v_i)}{\alpha} \cdot \ind{v_i \geq f(n, D)}} - \e[\vec v \leftarrow D]{\sum_{i = 1}^n \varphi(v_i) \cdot \ind{v_i > \beta_i(\vec v)}}
\end{align*}
The second line observes that if buyer $i$ receives the item, they pay at most $v_i$, and the auctioneer loses a collateral of $f(n, D)$ by withholding at least one bid. The third line invokes Lemma~\ref{lemma:alpha-regular}. To see that the assumptions for the Lemma are satisfied, let $E$ be the event where $v_i \geq f(n, D)$ and observe that $f(n, D) \geq r(D)$ for all $n \geq 1$ and $\alpha \in (0, 1)$. The fourth line observes $f(n, D) \geq r(D)$. The fifth line observes the event $\{\forall j, v_j < \beta_j(\vec v)\}$ implies $\{v_i < \beta_i(\vec v)\}$ and uses the fact $\beta_i(\vec v) > f(n, D)$. The sixth line uses the fact $\ind{a \leq X < b} = \ind{X \geq a} - \ind{X \geq b}$ for any random variable $X$ and constants $a > b$. The seventh line uses the fact $\alpha > 1$ and Observation~\ref{obs:single-candidate} which states the event $\{v_i > \beta_i\}$ implies $x_i(\vec v)$ and $v_j < \beta_j(\vec v)$ for all $j \neq i$ since $v_i$ expects to win the item. Moreover, we use the fact 
$$x_i(\vec v) \cdot \ind{\forall j \neq i, v_j < \beta_j(\vec v)} \cdot \ind{v_i \geq f(n, D)} \leq \ind{v_i \geq f(n, D)}.$$

To conclude, we must show that
\begin{align*}
    \e[\vec v \leftarrow D]{\sum_{i = 1}^n \frac{\varphi(v_i)}{\alpha} \cdot \ind{v_i \geq f(n, D)}} \leq \rev(D^n).
\end{align*}
From Lemma~\ref{lemma:single-revenue-upper-bound},
\begin{align*}
&\e[\vec v \leftarrow D]{\sum_{i = 1}^n \frac{\varphi(v_i)}{\alpha} \cdot \ind{v_i \geq f(n, D)}}\\
&\qquad= \frac{1}{\alpha}\left(\frac{1}{1-\alpha}\right)^{\frac{1}{1-\alpha}} \left(\frac{r(D)}{f(n, D)}\right)^{\frac{\alpha}{1-\alpha}}\e[\vec v \leftarrow D]{\sum_{i = 1}^n \varphi(v_i) \cdot \ind{v_i \geq r(D)}}\\ 
&\qquad\leq \frac{\alpha}{\alpha n}\e[\vec v \leftarrow D]{\sum_{i = 1}^n \varphi(v_i) \cdot \ind{v_i \geq r(D)}} \\
&\qquad = \frac{\alpha n}{\alpha n} \e[v_1 \leftarrow D]{\varphi(v_1) \cdot \ind{v_1 \geq r(D)}}\\
&\qquad = \rev(D)\\
&\qquad \leq \rev(D^n)
\end{align*}
The second line observes $f(n, D) \geq r(D)$ and applies Lemma~\ref{lemma:single-revenue-upper-bound}. The third line uses the assumption $f(n, D) \geq r(D) \left(\frac{n}{\alpha}\right)^{\frac{1-\alpha}{\alpha}}\left(\frac{1}{1-\alpha}\right)^{\frac{1}{\alpha}}$. The fourth line observes $\varphi(v_1), \ldots, \varphi(v_n)$ are i.i.d.. The fifth line observes $r(D)$ is the optimal reserve price, and so $\e[v_1 \leftarrow D]{\varphi(v_1) \cdot \ind{v_1 \geq r(D)}}$ is the optimal revenue for the single buyer auction (Theorem~\ref{thm:myerson}). The last line observes the revenue is non-decreasing in the number of buyers.
\end{proof}
Next, we prove our main result.
\begin{theorem}\label{thm:strongly-regular}
Assume the auctioneer follows a safe deviation to $\dra(f)$ with public broadcast and assume all buyer valuations are $\alpha$-strongly regular for $\alpha > 0$.
Then, there is an $f$ such that $\dra(f)$ with public broadcast is a credible auction.
\end{theorem}
\begin{proof}
Set $f(n, D) = r(D) \left(\frac{n}{\alpha}\right)^{\frac{1-\alpha}{\alpha}}\left(\frac{1}{1-\alpha}\right)^{\frac{1}{\alpha}}$. Observe for all $n \geq 1$ and $\alpha > 0$, $f(n, D) \geq r(D)$. For the case where $\alpha \geq 1$, the proof follows directly from Theorem~\ref{thm:ruduction-centralized-to-broadcast} because $f(n, D) \geq r(D)$. Next, consider the case where $\alpha \in (0, 1)$. Recall $\tilde b(n, D)$ refers to the highest bid among false buyers (or zero if no false buyer exists). $R(\vec v)$ refers to the auctioneer's revenue when buyers have value $\vec v$. For the case where $f(n, D) \geq \tilde b(n, D)$, Lemma~\ref{lemma:high-collateral} states the auctioneer's revenue is at most $\rev(D^n)$. Next, consider the case where $f(n, D) < \tilde b(n, D)$. We can write the revenue as
\begin{align*} &\e[\vec v \leftarrow D]{R(\vec v)} = \e[\vec v \leftarrow D]{R(\vec v) \cdot \ind{\exists j, v_j > \beta_j(\vec v)}} + \e[\vec v \leftarrow D]{R(\vec v) \cdot \ind{\forall j, v_j < \beta_j(\vec v)}}\\
&\qquad \leq \e[\vec v \leftarrow D]{\sum_{i = 1}^n \varphi(v_i) \cdot \ind{v_i > \beta_i(\vec v)}} + \rev(D^n) - \e[\vec v \leftarrow D]{\sum_{i = 1}^n \varphi(v_i) \cdot \ind{v_i > \beta_i(\vec v)}}\\
&\qquad = \rev(D^n)
\end{align*}
The second line is due to Lemma~\ref{lemma:small-collateral} and Lemma~\ref{lemma:safe-allocation}. This shows the auctioneer's revenue is at most $\rev(D^n)$ and proves there is a $f$ such that $\dra(f)$ is a credible auction.
\end{proof}

\section{Public Broadcast is Necessary}\label{sec:necessary}
This section revisits the fact centralized $\dra(f)$ is not a credible auction for certain $\alpha$-strongly regular valuations when $\alpha \in (0, 1)$.

\begin{theorem}[Theorem 4.4 in \cite{ferreira2020credible}]\label{thm:centralized-dra-negative} For all $f$, $\alpha \in (0, 1)$, there exists a $D^n$ that is $\alpha$-strongly regular such that centralized $\dra(f)$ is not credible for instance $D^n$.
\end{theorem}

The following is a special case for the instance given in the proof of Theorem~\ref{thm:centralized-dra-negative}.
By inspection, this strategy is a safe deviation for centralized $\dra(f)$ since, in the view of each buyer, the strategy is indistinguishable from the promised auction.
In this strategy the auctioneer only sends a shill bids to buyer $B$ that depend on the bid of buyer $A$.
This would not be possible if, rather than relying on the auctioneer to forward messages, messages were sent in a broadcast channel because any message one buyer receives is also received by other buyers.

\begin{definition}[Adaptive Reserve Price]\label{def:adaptive}
Consider an auctioneer who promises to implement centralized $\dra(f)$ on an instance with two buyers $A$ and $B$. The {\em adaptive reserve price deviation} is the following deviation:
\begin{itemize}

\item $A$ sends $(A, c_A)$ to the auctioneer.

\item $B$ sends $(B, c_B)$ to the auctioneer.

\item The auctioneer sends $(B, c_B)$ to $A$ and $(A, c_A)$ to $B$

\item The auctioneer sends ``End of the Commitment Phase'' to buyer $A$, then requests $A$ to reveal their bid.
$A$ complies and reveals $(b_A, r_A)$ such that $c_A = \commit(b_A, r_A)$.

\item The auctioneer picks a large threshold $T$:

\begin{itemize}
\item If $b_A < T$, the auctioneer sends ``End of Commitment Phase'' to buyer $B$, then requests $B$ to reveal their bid (who complies by revealing $(b_B, r_B)$ such that $c_B = \commit(b_B, r_B)$).
The auctioneer implements the allocation/payment rule for the second-price auction with reserve $r(D)$ on bids $\{b_A, b_B\}$.

\item If $b_A \geq T$, the auctioneer impersonates a false buyer $C$.
Let $r_C$ be uniformly random and $b_C = b_A + f(2, D)$.
Then, the auctioneer sends $(C, \commit(b_C, r_C))$ to $B$.
The auctioneer sends ``End of Commitment Phase'' to buyer $B$, then request $B$ to reveal their bid.
$B$ complies and reveal $(b_B, r_B)$ such that $c_B = \commit(b_B, r_B)$.
Next, the auctioneer proceeds as follows:

\begin{itemize}
	\item If $r(D) \geq \max\{b_A, b_B\}$, the auctioneer reveals all bids. No one receives the item.
	
    \item If $b_B < b_A$ and $b_A > r(D)$, the auctioneer reveals all bids and allocates the item to $A$ and charges $\max\{r(D), b_B\}$.
    
    \item If $b_B \in [b_A, b_C]$ and $b_B > r(D)$, the auctioneer reveals $b_A$ and hides $b_C$ from $B$. Then, the auctioneer allocates the item to $B$ and charges $\max\{b_A, r(D)\}$.
    
    \item If $b_B > b_C$, the auctioneer reveals all bids and allocates the item to $B$ who pays $b_C$.
    \end{itemize}
\end{itemize}
\end{itemize}
\end{definition}

\section{DRA over Public Broadcast for Regular Distributions}\label{sec:regular}
Although $\dra$ with public broadcast extends the class of distributions where it is credible, it is not a magic bullet.
Indeed, Theorem~\ref{thm:regular} states there is an instance with a single buyer drawn from a regular distribution that witnesses $\dra(f)$ with public broadcast is not credible.
The proof relies on a similar negative result in \cite{ferreira2020credible}. 

\begin{theorem}[Theorem 4.4 in \cite{ferreira2020credible}]\label{thm:centralized-dra-regular}
There is a regular distribution $D$ such that for all $f(\cdot, \cdot)$, centralized $\dra(f)$ is not credible even when there is a single buyer with valuation drawn from $D$.
\end{theorem}

\begin{theorem}\label{thm:regular}
There is a regular distribution $D$ such that for all $f(\cdot, \cdot)$, $\dra(f)$ over public broadcast is not credible even when there is a single buyer with a valuation drawn from $D$.
\end{theorem}
\begin{proof}
We will argue for any instance with a single buyer, any safe deviation to centralized $\dra(f)$ maps to a safe deviation to $\dra(f)$ over public broadcast.
To see, let $(G, \vec s)$ be a safe deviation to centralized $\dra(f)$. Let $(G', \vec s')$ be a deviation to $\dra(f)$ with public broadcast identical to $(G, \vec s)$ except on the following cases:
\begin{itemize}
    \item Whenever the buyer sends $m$ to the auctioneer in $(G, \vec s)$, the buyer broadcast $m$ in $(G', \vec s')$.

    \item Whenever the auctioneer sends $m$ to the buyer in $(G, \vec s)$, the auctioneer broadcast $m$ in $(G', \vec s')$.
\end{itemize}
$(G', \vec s')$ is a safe deviation because $(G, \vec s)$ is a safe deviation.
Moreover, $(G', \vec s')$ induces the same allocation/payment rule as $(G, \vec s)$; therefore, $\rev(G', \vec s') = \rev(G, \vec s)$.

From Theorem~\ref{thm:centralized-dra-regular}, there is a $D$ such that for all $f(\cdot, \cdot)$, there is a safe deviation $(G, \vec s)$ to centralized $\dra(f)$ where $\rev(G, \vec s) > \rev(D)$.
The mapping above proves there is a safe deviation $(G', \vec s')$ to $\dra(f)$ with public broadcast where $\rev(G', \vec s') = \rev(G, \vec s) > \rev(D)$.
This proves $\dra(f)$ with public broadcast is not a credible auction on instance $D$ as desired.
\end{proof}

\section{Conclusion}\label{sec:conclusion}
Improving the transparency and fairness in Internet platforms is becoming an essential concern for regulators, as observed by the US Department of Justice lawsuit against Google~\cite{doj2023}.
It is unlikely that customers could unilaterally detect and, more importantly, prove the sophisticated market manipulations alleged in the complaint.
Credible auctions formalize the notion that an auction is ``auditable'' by its participants: the auctioneer has no incentive to deviate from running the promised mechanism in earnest.
However, existing credible auctions suffer from restrictive assumptions on valuation distributions and exclude valuations with tails thicker than the exponential distribution.

This work shows that censorship-resistant broadcast channels like blockchains are helpful to circumvent this problem. 
We propose the deferred revelation auction with public broadcast, a natural modification of the centralized deferred revelation auction of \cite{ferreira2020credible}.
Although our auction represents a simple modification of a known auction, the resulting auction is credible in instances where no known communication-efficient auctions were known to be credible.
This work builds on the emerging line of research that attempts to improve the performance of economic mechanisms by appending cryptographic primitives to them.
The need for large collateral is a limitation of our work.
Minimizing collateral is an important objective to make these auctions practical which we leave as future direction.

\bibliographystyle{acm}
\bibliography{ms}

\appendix

\section{Mathematical Background}\label{app:background}
\begin{lemma}[Lemma 7.1 in \cite{ferreira2020credible}]\label{lemma:alpha-regular}
Let $D$ be $\alpha$-strongly regular for $\alpha > 0$. Let $E$ be an event such that $v \geq r(D)$ with probability $1$ conditioned on $E$. Then
$$\e{v | E} \leq \frac{1}{\alpha}\e{\varphi^D(v) | E} + r(D).$$
\end{lemma}
\begin{proof}
Because $D$ is $\alpha$-strongly regular, for all $x' > x$,
$$\varphi^D(x') - \varphi^D(x) \geq \alpha(x' - x)$$

Then for any $x' \geq r(D)$, $x' \leq \frac{1}{\alpha}(\varphi^D(v) - \varphi^D(r(D))) + r(D)$. By definition $\varphi^D(r(D)) = 0$. Conditioned on event $E$, we have that $v \geq r(D)$ for all $v$. We conclude $\e[\vec v \leftarrow D]{v | E} \leq \frac{1}{\alpha} \e{\varphi^D(v) | E} + r(D)$ as desired.
\end{proof}

\begin{lemma}[Lemma 7.2 in \cite{ferreira2020credible}]\label{lemma:alpha-tail}
    Let $D$ be a $\alpha$-strongly regular distribution. Then for all $p \geq r(D)$,
    $$p \cdot \pr[\vec v \leftarrow D]{v \geq p} \leq r(D) \cdot \pr[\vec v \leftarrow D_0]{v \geq r(D)} \left(\frac{1}{1-\alpha}\right)^{\frac{1}{1-\alpha}} \left(\frac{r}{p}\right)^{\frac{\alpha}{1-\alpha}}.$$
\end{lemma}

\begin{lemma}\label{lemma:single-revenue-upper-bound}
Let $D$ be a $\alpha$-strongly regular for $\alpha > 0$. Then for all $p \geq r(D)$,
$$\e[\vec v \leftarrow D]{\varphi(v) \cdot \ind{v \geq p}} \leq \e[\vec v \leftarrow D]{\varphi(v) \cdot \ind{v \geq r(D)}}\left(\frac{1}{1-\alpha}\right)^{\frac{1}{1-\alpha}} \left(\frac{r}{p}\right)^{\frac{\alpha}{1-\alpha}}.$$
\end{lemma}
\begin{proof}
Consider a single item, single bidder posted-price mechanism that offers the item at a price $p$. The bidder value is drawn from $D$. The revenue is $p \pr[\vec v \leftarrow D]{v \geq p}$ because the buyer purchases whenever their value exceeds $p$. From Myerson's theorem, $p \pr[\vec v \leftarrow D]{v \geq p} = \e[\vec v \leftarrow D]{\varphi(v) \cdot \ind{v \geq p}}$. The result follows directly by applying Lemma~\ref{lemma:alpha-tail} to the left-hand side of the inequality.
\end{proof}



\end{document}